\documentclass[english,aps,amssymb,prl,twocolumn, showpacs]{revtex4-1}
\pdfoutput=1
\usepackage[T1]{fontenc}
\usepackage[latin9]{inputenc}
\usepackage{amsmath}
\usepackage{graphicx}
\usepackage{amssymb}
\usepackage{esint}

\makeatletter
\@ifundefined{textcolor}{}
{
 \definecolor{WHITE}{gray}{1}
 \definecolor{RED}{rgb}{1,0,0}
 \definecolor{GREEN}{rgb}{0,1,0}
 \definecolor{BLUE}{rgb}{0,0,1}
 \definecolor{CYAN}{cmyk}{1,0,0,0}
 \definecolor{MAGENTA}{cmyk}{0,1,0,0}
 \definecolor{YELLOW}{cmyk}{0,0,1,0}
 }

\renewcommand{\phi}{\varphi}
\renewcommand{\epsilon}{\varepsilon}

\renewcommand{\vec}[1]{{\bf #1}}
\newcommand{Û}{$}

\usepackage{babel}

\begin{document}

\title {Tuning topological superconductivity in helical Shiba chains by supercurrent}
\author{Joel R\"ontynen}
\author{Teemu Ojanen}
\email[Correspondence to ]{teemuo@boojum.hut.fi}
\affiliation{O. V. Lounasmaa Laboratory (LTL), Aalto University, P.~O.~Box 15100,
FI-00076 AALTO, Finland }
\date{\today}
\begin{abstract}
Recent experimental investigations of arrays of magnetic atoms deposited on top of a superconductor have opened a new chapter in the search of topological superconductivity.  We generalize the microscopic model derived by Pientka et al. [Phys. Rev. B \textbf{88}, 155420 (2013)] to accommodate the effects of finite supercurrent in the host material. Previously it was discovered that helical chains with nonplanar textures are plagued by a gapless phase. We show that by employing supercurrent it is possible to tune the chain from the gapless phase to the topological gapped phase. It is also possible to tune the chain between the trivial and the topological gapped phase, the size of which may be dramatically increased due to supercurrent. For planar textures supercurrent mainly contributes to proliferation of the gapless phase. Our predictions, which can be probed in STM experiments, are encouraging for observation and manipulation of Majorana states.

\end{abstract}
\maketitle
\bigskip{}

\emph{Introduction}-- Finding novel realizations for topological superconductivity and accompanying Majorana states has become a major source of inspiration in quantum condensed matter physics\cite{qi, schnyder, kitaev1}. The possibility of engineering Majorana bound states, particle-like entities that could serve as building blocks of topological quantum computation \cite{nayak, kitaev2, kitaev3, alicea2}, has been the primary driving force in the recent developments. Magnetic Shiba chains \cite{choy,np, vazifeh, poyh, pientka2, pientka3, heimes}, consisting of arrays of magnetic atoms deposited on top of a superconducting host material,  are recent promising candidates for topological superconductivity and have just become accessible in experiments \cite{NP}.  Magnetic realizations of topological superconductivity have attracted attention, since they commonly circumvent the need for materials with strong spin-orbit coupling or exotic superconducting pairing \cite{braunecker1, klinovaja2, klinovaja, kjaergaard, ojanen2, klinovaja3, braun}. Shiba chains are particular representatives of magnetic systems consisting of magnetic atoms on top of a host s-wave superconductor. Nearby magnetic atoms form effectively one-dimensional (1d) band that may undergo a  topological phase transition to 1d topological superconductor with Majorana end states similar to nanowire realization \cite{lutchyn, oreg, mourik, das}. In Shiba chains are exceptionally disorder free and allow accessing the local density-of-state (LDOS) in Scanning Tunneling Microscope (STM) experiments, enabling spatial mapping of the Majorana wavefunctions.

Previous work on Shiba chains mostly employ a short-range hopping model which obeys the correct symmetries and captures some qualitative features of the topological properties \cite{choy,np, vazifeh, poyh}. However, this model cannot describe the physically most relevant case where the superconducting coherence length is much longer than the separation between adjacent magnetic atoms. A substantial step was taken by Pientka \emph{et al.}, who provided a microscopic derivation of a long-range hopping model \cite{pientka2, pientka3} for helical magnetic order \cite{menzel}, arising possibly from the RKKY and the spin-orbit interaction. This model allows analytical studies of the quantitative features of the topological phase diagram in the physically relevant regime.    

In this work we follow the treatment of Ref.~\cite{pientka2} and generalize the microscopic theory of helical Shiba chains to the case where the order parameter of the host superconductor supports supercurrent. Previously it was discovered that the phase diagram of a non-planar helical chain exhibits a pervasive gapless phase \cite{pientka2}. Remarkably, we discover that supercurrent enables tuning a gapless system to the topological gapped phase. This is an important difference compared to the nanowire realization where the system is always gapped in the absence of supercurrent \cite{romito}. For a non-planar helix it is possible to push the system into a topological state even if the topological phase is marginal or completely \emph{absent} without  supercurrent and to switch the state between the topological and trivial phases. This is very encouraging for observing topological superconductivity manipulating Majorana states. When the magnetic helix is planar, supercurrent will mainly drive the system towards the gapless phase.  In contrast to Ref.~\cite{heimes} which studied the effects of supercurrent on antiferromagnetic chains (a special point in the space of helical textures),  our results yield an essentially analytical description for the phase diagram of \emph{arbitrary} helical textures, allowing a simple physical interpretation of the supercurrent effects. We will also provide a direct connection of our theory of supercurrent induced control to observables by calculating the LDOS which can be accessed in STM experiments.               

\emph{Model}-- We consider a superconducting system with a regular 1d lattice of magnetic atoms deposited on top of it. Our strategy follows the formulation of Ref.~\cite{pientka2} which is complicated by the supercurrent-induced modifications. Working in the Nambu operator basis $\hat{\Psi}=(\hat{\psi}_{\uparrow},\hat{\psi}_{\downarrow},\hat{\psi}_{\downarrow}^\dagger,-\hat{\psi}_{\uparrow}^\dagger)^T$, the Bogoliubov-de Gennes equation for the four component c-number spinor becomes
\begin{equation} \label{h1}
[E-\xi_k\tau_z-\vec{\Delta}\cdot\tau]\Psi(\vec{r})=-J\sum_{j}\vec{S}_j\cdot \sigma\delta(\vec{r}_j)\Psi(\vec{r}),
\end{equation}
where $\vec{\Delta}=|\Delta|\left(\cos\phi(r), \sin\phi(r),0\right)$ describes the position-dependent superconducting order parameter, $\xi_k=\frac{k^2}{2m}-\mu$ is the single-particle energy, $J$ is the exchange coupling  and $\vec{S}_j$ describes the direction and magnitude of the magnetic moment of $j$th atom. Supercurrent flowing in the bulk is proportional to $\nabla \phi$ The set of Pauli matrices $\vec{\tau}=(\tau_x,\tau_y, \tau_z)$ and $\vec{\sigma}=(\sigma_x,\sigma_y, \sigma_z)$ operate in particle-hole and spin space, respectively. In this work we consider magnetic textures of the form $\hat{\vec{S}}_j=(\cos 2k_haj \sin \theta, \sin 2k_haj \sin \theta, \cos \theta)$, where $k_h$ is the wave number of the magnetic helix pitch angle, $\theta$ is the tilt of the moments and $a$ is the distance between two adjacent moments. A helix is planar when $\theta=\pi/2$ in which case the average magnetization vanishes.  

The explicit breaking of translational invariance on the LHS in (\ref{h1}) due to the superconducting phase poses a major complication in solving the eigenvalue problem. To remedy this, we introduce unitary transformation $\bar\Psi(\vec{r})=U(r)\Psi(\vec{r})$ where $U(r)=e^{i\tau_z\frac{\phi(r)}{2}}$. The transformed Eq.~(\ref{h1}) takes the form
\begin{align}\label{h2}
[E-\tau_z&\left(\frac{(\vec{k}-\frac{\nabla\phi}{2}\tau_z)^2}{2m}-\mu\right)-|\Delta|\tau_x]\bar{\Psi}(\vec{r})\nonumber\\
=&-J\sum_j\vec{S}_j\cdot \sigma\delta(\vec{r}_j)\bar{\Psi}(\vec{r}).
\end{align}
We will further assume that the phase winding (and thus the supercurrent) is linear so that $\nabla\phi$ is independent of $\vec{r}$. The key observation here is that in the transformed basis the explicit violation of the translation symmetry on the LHS has been removed,  allowing us to follow the general steps of Ref.~\cite{pientka2} with some additional technical complications.  The equation for the spinor at the position of $i$th atom can then be written as
\begin{equation} \label{psi3}
\bar{\Psi}(\vec{r}_i)=-\sum_{j}(\vec{\hat{S}}_j\cdot \sigma)J_E(\vec{r}_i-\vec{r}_j)\bar{\Psi}(\vec{r}_j),
\end{equation}
where $\vec{\hat{S}}_j=\vec{S}_j/S$, $S=|\vec{S}_j|$ and 
\begin{align} \label{h3}
&J_E(\vec{r})=JS\int\frac{d\vec{k}}{(2\pi)^3} \frac{e^{i\vec{k}\cdot \vec{r}}}{E-\tau_z\left(\frac{(\vec{k}-\frac{\nabla\phi}{2}\tau_z)^2}{2m}-\mu\right)-|\Delta|\tau_x}
\nonumber\\
&=JS\int\frac{d\vec{k}}{(2\pi)^3}e^{i\vec{k}\cdot \vec{r}} \frac{(E+\frac{\vec{k}\cdot\nabla\phi}{2m})+\tau_z\xi_k+|\Delta|\tau_x}{(E+\frac{\vec{k}\cdot\nabla\phi}{2m})^2-\xi_k^2-|\Delta|^2}
\end{align}
On the second line we have noted that the characteristic  magnitude of momentum is $|\vec{k}|\approx k_F$ and the maximum phase gradient corresponding to the critical current satisfies $|\nabla\phi|\lesssim 2\pi/\xi\ll k_F$, where the coherence length is defined as $\xi=v_F/|\Delta|$, so that the $(\nabla\phi)^2$ term in the denominator gives a negligible contribution to the term proportional to $\tau_z$. From Eq.~(\ref{h3}) we see that the phase gradient introduces a new energy scale $\tilde{\epsilon}_\phi=\frac{v_F|\nabla\phi|}{2}$ in the problem  which satisfies condition $\tilde{\epsilon}_\phi/|\Delta|\ll1$ when the supercurrent is much smaller than the critical current.  Since this is the relevant regime for finding robust gapped states in general, we can treat $\tilde{\epsilon}_\phi/|\Delta|$ as a natural small parameter.

Following Ref.~\cite{pientka2} we first consider the case of a single magnetic moment to understand the low-energy properties of a chain. This problem can be solved starting from Eq.~(\ref{psi3}) by setting $\vec{S}_j=0$ when $j\neq i$ and evaluating $J_E(0)$ \cite{on}. In the absence of supercurrent ($\tilde{\epsilon}_\phi=0$)  there exist two sub-gap solutions $E=\pm \epsilon=\pm|\Delta|\frac{1-\alpha^2}{1+\alpha^2}$ where $\alpha=\pi\nu_0SJ$ is a dimensionless constant \cite{yu, shiba, rusinov}.  The corresponding eigenspinors are $\Psi_+(\vec{r}_i)\equiv|+i\rangle=|+\tau_x\rangle|\uparrow i\rangle$ and $\Psi_-(\vec{r}_i)\equiv|-i\rangle=|-\tau_x\rangle|\downarrow i\rangle$, where $\tau_x|\pm\tau_x\rangle=\pm|\pm\tau_x\rangle$ and  $\vec{\hat{S}}_i\cdot\sigma| \uparrow/ \downarrow i\rangle=\pm|\uparrow/ \downarrow i\rangle$.  For deep impurities defined by $\alpha\sim 1$ that we study below, the bound-state energies are $\pm\epsilon_0=\pm|\Delta|(1-\alpha)$. The  supercurrent-induced modification in the lowest order is $\pm\tilde{\epsilon}_0 =  \pm\epsilon_0\mp \frac{|\Delta|}{6}\left(\frac{\tilde{\epsilon}_\phi}{|\Delta|} \right)^2$ while leaving the eigenstates unaffected \cite{on}.

The single-impurity sub-gap states form a convenient basis to study the low-energy properties of Shiba chains. Expanding the spinors $\bar{\Psi}(\vec{r}_i)$ in the low-energy components $\Psi'_i=\left(\langle +i|\bar{\Psi}(\vec{r}_i)\rangle, \langle -i|\bar{\Psi}(\vec{r}_i)\rangle\right)^T$ and projecting Eq.~(\ref{psi3}) onto the low-energy subspace as in Ref.~\cite{pientka2}, we discover an
effective Bogoliubov-de Gennes equation for the reduced spinor $H\Psi'=E\Psi'$, where 
\begin{equation} \label{Hm}
H=
\left(
\begin{array}{cc}
  h_{ij}   &\Delta_{ij}   \\
  (\Delta_{ij})^\dagger   &-h^*_{ij}   \\ 
\end{array}
\right).
\end{equation}
Here $h_{ij}=\tilde{\epsilon}_0$ and $\Delta_{ij}=0$ when $i=j$ and
\begin{align}\label{hij2} 
h_{ij}=&|\Delta|\frac{e^{-\frac{r_{ij} }{\xi}} }{k_Fr_{ij}}\left(i\frac{\epsilon_\phi}{|\Delta|}\,\text{sign}(i-j)\cos{k_Fr_{ij}}-\sin{k_Fr_{ij}} \right) \langle \uparrow i|\uparrow j\rangle\nonumber\\
\Delta_{ij}=&|\Delta|\frac{e^{-\frac{r_{ij} }{\xi}} }{k_Fr_{ij}}\cos{k_Fr_{ij}}\langle \uparrow i|\downarrow j\rangle,
\end{align}
when $i\neq j$, with $r_{ij}=|\vec{r}_i-\vec{r}_j|=a|i-j|$ \cite{on}. The parameter $\epsilon_\phi=\frac{v_F|\nabla\phi|}{2} \cos \beta $, where $\beta$ is the angle between $\nabla\phi$ and $\vec{r}_i-\vec{r}_j$, describes the magnitude and the direction of the supercurrent. Model (\ref{Hm}) is valid for the case of deep impurities $\epsilon_0/|\Delta|\ll1$ in the dilute limit $k_Fa\gg1$. When $\epsilon_\phi=0$,  Eq.~(\ref{hij2}) reduces to the result derived in Ref.~\cite{pientka2}.  The supercurrent-dependent corrections to Eq.~(\ref{hij2}) are of the order $\mathcal{O}(\frac{1}{k_Fa}\frac{\epsilon_0\epsilon_\phi}{|\Delta|^2}, \frac{1}{k_Fa}\frac{\epsilon_\phi^2}{|\Delta|^2})$ containing three small parameters and are omitted below. The spin matrix elements corresponding to the helical order are  $\langle \uparrow i|\uparrow j\rangle=\cos^2{\frac{\theta}{2}}+\sin^2{\frac{\theta}{2}}e^{-2ik_hx_{ij}}$ and $\langle \uparrow i|\downarrow j\rangle=i\sin{k_hx_{ij}}\sin{\theta}$, where $x_{ij}=a(i-j)$. To benchmark our result with that of Ref.~\cite{pientka2}, we also perform a unitary transformation modifying the matrix elements $\langle \uparrow i|\uparrow j\rangle\to\cos^2{\frac{\theta}{2}}e^{i k_hx_{ij}} +\sin^2{\frac{\theta}{2}}e^{-ik_hx_{ij}}$ and $\langle \uparrow i|\downarrow j\rangle=i\sin{k_hx_{ij}}\sin{\theta}$.

The Hamiltonian (\ref{Hm}) is appropriate for finite-size studies but the topological phase diagram is most conveniently studied in Fourier space. Evaluating $h(k)=\sum_{j}h_{ij}e^{ikx_{ij}}$  and $ \Delta(k)$ we obtain 
\begin{equation} \label{Hk}
H(k)=
\left(
\begin{array}{cc}
  h(k)   &\Delta(k)   \\
  \Delta(k)   &-h(-k)   \\ 
\end{array}
\right)
\end{equation}
where $h(k)$ can be expressed through the antisymmetric and symmetric components
$\mathcal{A}h(k)=\frac{1}{2}(h(k)-h(-k))$ and $\mathcal{S}h(k)=\frac{1}{2}(h(k)+h(-k))$ as
\begin{align}
&\mathcal{A}h(k)=-\frac{|\Delta|}{k_Fa}\frac{\cos{\theta}}{2} \left[ f(k_1)+f(k_2)-f(k_3)-f(k_4)  \right] -\nonumber\\
&-\frac{1}{2}\frac{\epsilon_\phi}{k_Fa} \left[ f(k_1)+f(k_2)+f(k_3)+f(k_4)  \right], \label{ahk} \\
&\mathcal{S}h(k)=\tilde{\epsilon}_0-\frac{|\Delta|}{k_Fa}\frac{1}{2} \left[ f(k_1)-f(k_2)+f(k_3)-f(k_4)  \right] -\nonumber\\
&-\frac{\epsilon_\phi}{k_Fa}\frac{\cos{\theta}}{2} \left[ f(k_1)-f(k_2)-f(k_3)+f(k_4)  \right], \label{shk}
\end{align}
in terms of the variables $k_1=k+k_F+k_h$, $k_2=k-k_F-k_h$, $k_3=k+k_F-k_h$, $k_4=k-k_F+k_h$ and the function $f(k)=\text{arctan}\frac{e^{-a/\xi}\sin ka}{1-e^{-a/\xi}\cos ka}$. The (antisymmetric) effective pairing function is given by   
\begin{align}\label{Dk}
&\Delta(k)=\frac{|\Delta|}{k_Fa}\frac{\sin{\theta}}{4} \left[ \tilde{f}(k_1)-\tilde{f}(k_2)-\tilde{f}(k_3)+\tilde{f}(k_4)  \right]
\end{align}
where $\tilde{f}(k)=-\text{ln}\left(1+e^{-2a/\xi}-2e^{-a/\xi}\cos ka \right)$.
The spectrum of Eq.~(\ref{Hk}) is 
\begin{align}\label{E}
E_{\pm}(k)=\mathcal{A}h(k)\pm \sqrt{ (\mathcal{S}h(k))^2+\Delta(k)^2 }.
\end{align}

\emph{Supercurrent-modified phase diagram}--  The analytically obtained  spectrum $(\ref{E})$ allows us to make a number of general statements.  A finite antisymmetric function $\mathcal{A}h(k)$ always suppresses the gap between the $E_+$ and $E_-$ bands compared to the case $\mathcal{A}h(k)=0$. Since the system obeys  particle-hole symmetry $E_+(k)=-E_-(-k)$, we immediately see that the system enters a gapless phase whenever min $E_+(k)<0$, a transition driven by $\mathcal{A}h(k)$.  The phase transition between topological and trivial gapped phases requires that the square root in Eq.~$(\ref{E})$ vanishes, taking place when $\mathcal{S}h(k)=0$ and $\Delta(k)=0$, while $E_+(k)\geq 0$ should be satisfied. This can only take place at $k_0a=n\pi$ ($n=0,\,\pm1...$) where $\Delta(k_0)=0$. In summary, $\mathcal{A}h(k)$ may only drive the gapless-gapped transitions and $\mathcal{S}h(k)$ only affects the transitions between the gapped phases. These general properties have crucial implications for the supercurrent-modified phase diagram in different cases of interest.

\begin{figure}
\includegraphics[width=0.9\columnwidth]{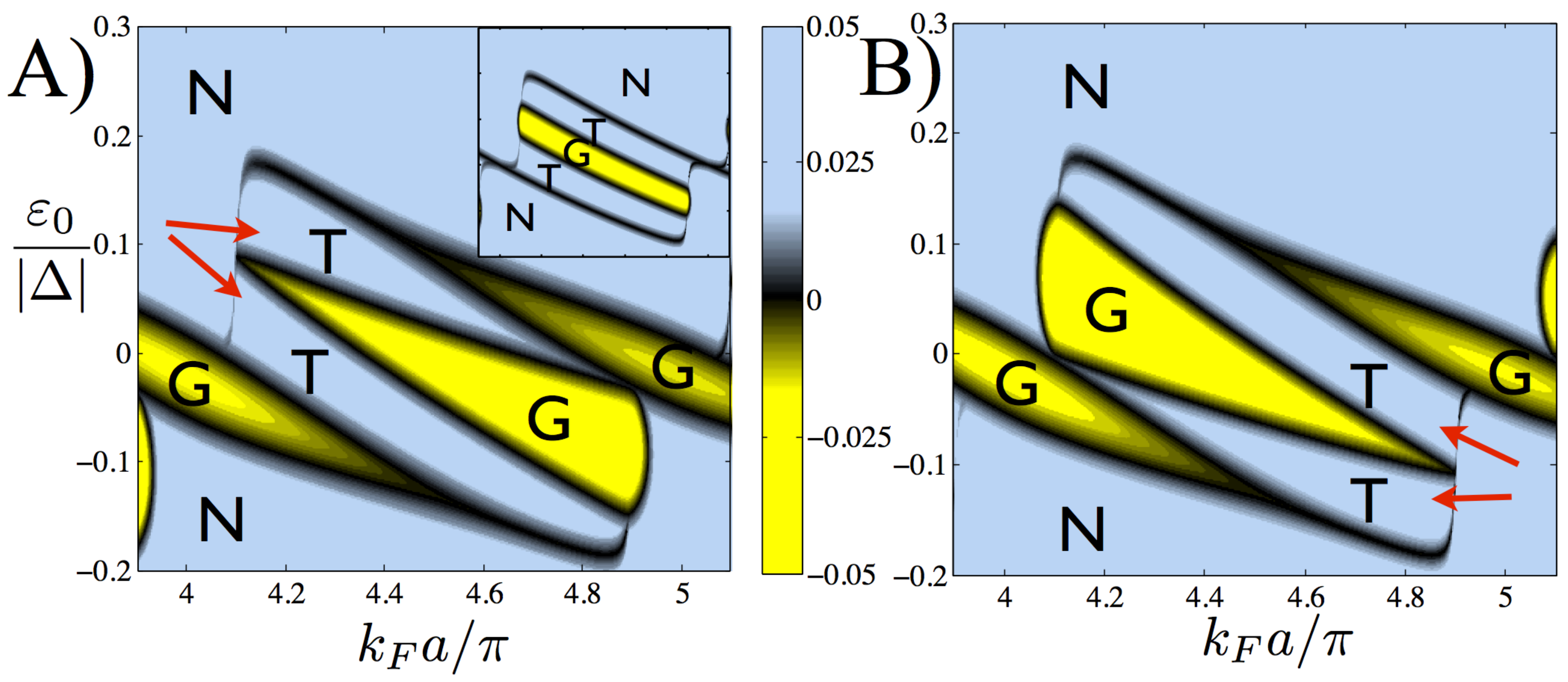}
\caption{ A): Minimum value of $E_+(k)$. Different phases are separated by the condition min $E_+(k)=0$. The labels stand for normal (N), topological (T) and gapless (G). The parameters are $\theta=2\pi/5, k_ha=\pi/10, \epsilon_\phi=|\Delta|/3, \xi=50a$. The inset shows the phase diagram for vanishing supercurrent $\epsilon_\phi=0$. In addition to adding gapless regions, finite supercurrent also pushes some gapless regions to the topologically nontrivial gapped phase indicated by the red arrows. B): Same as A), but with inverted  supercurrent $\epsilon_\phi=-\epsilon_\phi$. }\label{fig1}
\end{figure}

Case $1^\circ$: The helical texture is nonplanar $\theta\neq \frac{\pi}{2}$ and the parameters put the system in the gapless phase in the absence of supercurrent.  The gapless phase arises from finite $\mathcal{A}h(k)$ which dominates the square root in $(\ref{E})$ for some $k$. By reducing  $\mathcal{A}h(k)$ by the supercurrent-induced contribution in the second line of Eq.~$(\ref{ahk})$, it is possible to drive the system towards a gapped state. As illustrated in Fig.~\ref{fig1}, the supercurrent-induced gapped state can be the topologically nontrivial so it is possible to tune the system from the gapless initial state to the topologically nontrivial gapped state.   

\begin{figure}
\includegraphics[width=0.9\columnwidth]{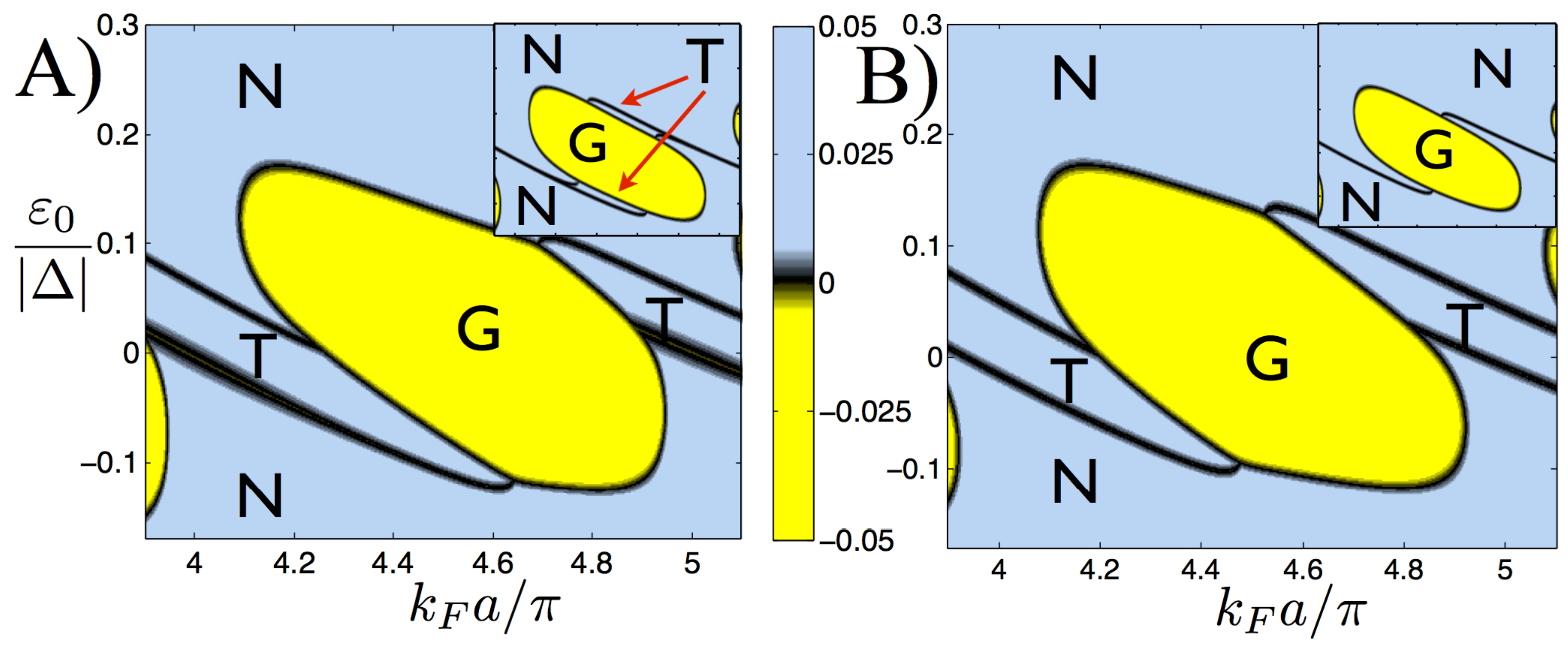}
\caption{ A): Same quantities as in Fig.~\ref{fig1} but for $\theta=\pi/5, k_ha=\pi/3, \epsilon_\phi=|\Delta|/3, \xi=50a$. The inset shows the phase diagram for vanishing supercurrent $\epsilon_\phi=0$. Supercurrent significantly increases the size of the topological region.  B): Same as A), but $k_ha=\frac{\pi}{2}$. Supercurrent opens up large topologically nontrivial regions that are completely absent when the supercurrent vanishes. }\label{fig2}
\end{figure}

Case $2^\circ$: The helical texture is nonplanar $\theta\neq \frac{\pi}{2}$ and the parameters put the system in a gapped phase in the absence of supercurrent. The second line of Eq.~(\ref{shk}) indicates that the supercurrent-induced term modifies the symmetric function $\mathcal{S}h(k)$ which drives the transition between the topological and trivial gapped phases. This  mechanism enables supercurrent-induced switching between the gapped phases. This could take place in the regions of the phase diagram where the condition $\mathcal{S}h(k_0)=0$ is met before the supercurrent-induced contribution to  $\mathcal{A}h(k)$ drives the system gapless.  Remarkably, as illustrated in Fig.~\ref{fig2} A), for some helical configurations it is possible to dramatically increase the topologically nontrivial region in the phase diagram. Ultimately, as depicted in Fig.~\ref{fig2} B), it is also possible to open up topological regions in the antiferromagnetic case $k_ha=\pi/2$ where they are completely absent for vanishing supercurrent \cite{heimes}.   

\begin{figure}
\includegraphics[width=0.99\columnwidth]{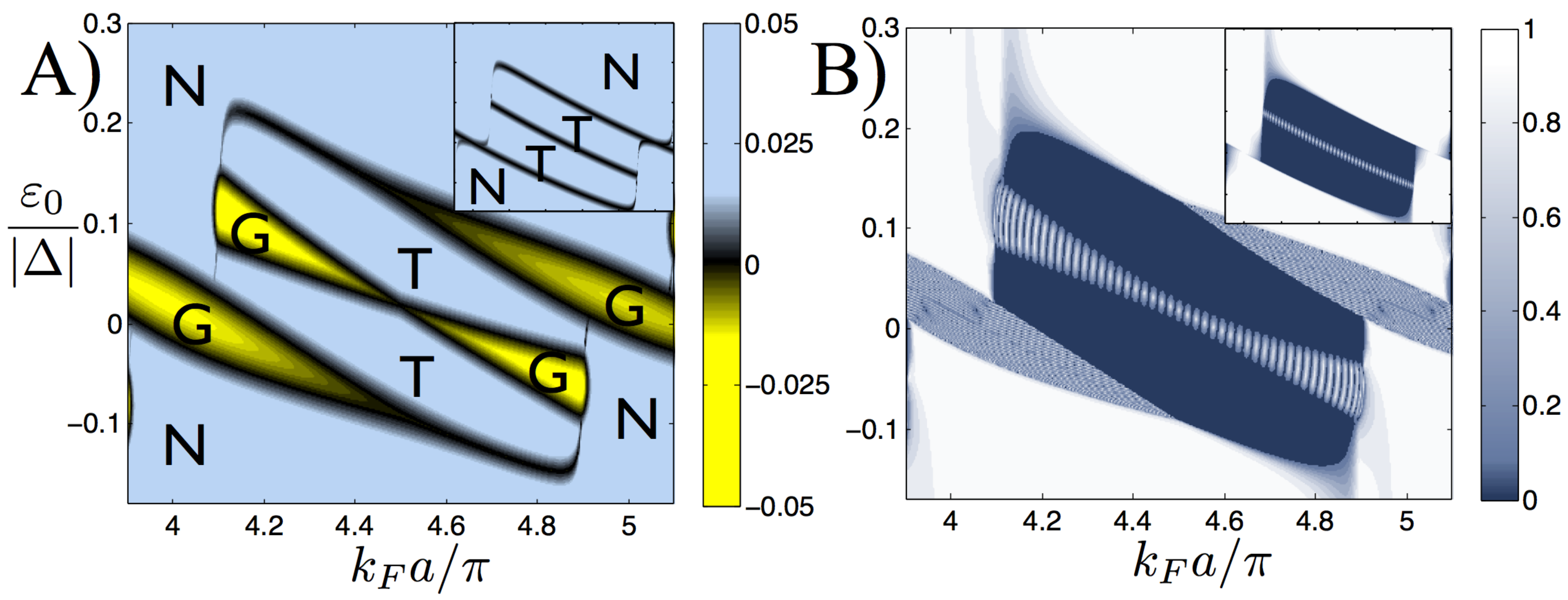}
\caption{A): Same quantities as in Fig.~\ref{fig1} but for a planar helix $\theta=\frac{\pi}{2}, k_ha=\pi/10, \epsilon_\phi= |\Delta|/3, \xi=50a$. Supercurrent cannot deform the phase boundaries between the gapped phases, it can only add gapless regions on top of $\epsilon_\phi=0$ phase diagram (in the inset).  B): Phase diagram for a finite-size chain of 50 atoms with the same parameters as in A). Colors indicate the ratio of the lowest-lying and the first excited energy level $|E_0/E_1|$. Small values (dark blue) indicate the topological phase with Majorana end states and large values (white) signals the trivial gapped state. In a finite-size system oscillations (blue stripes) indicate the gapless phase. }\label{fig3}
\end{figure}

Case $3^\circ$: The helical texture is planar $\theta= \frac{\pi}{2}$. In the absence of supercurrent the phase diagram contains only gapped phases (for $k_ha\neq n\pi$) \cite{pientka2}.  The linear supercurrent-induced term on the second line of Eq.~(\ref{shk}) vanishes, so the supercurrent modification to $\mathcal{S}h(k)$ arises from the weak supercurrent renormalization of $\tilde{\epsilon}_0$. However, the antisymmetric contribution, which is zero for planar texture in the absence of supercurrent, becomes nonzero due to the second line of Eq.~(\ref{ahk}). As pointed out above, finite $\mathcal{A}h(k)$ only have detrimental effect on the gapped phases, suppressing gaps and eventually driving the system to the gapless phase. As illustrated in Fig.~\ref{fig3} A), supercurrent mainly adds a gapless region in the phase diagram but does not deform the phase boundaries between the gapped phases. Numerical diagonalization of finite-size systems shows that the phase diagram of an infinite system is reproduced accurately with a few tens of magnetic sites as shown in Fig.~\ref{fig3} B) and essentially perfectly with $\gtrsim 100$ lattice sites. 

\emph{Physical implications}-- Experimental investigation of topological properties of Shiba chains has recently been initiated. In STM experiments it is possible to map the LDOS $N_i(E)=\sum_n \left[ |u_n(i)|^2\delta(E-E_n)+ |v_n(i)|^2\delta(E+E_n)\right]$ along the chain. Here $u_n(i)$ ($v_n(i)$) is the particle-like (hole-like) component of the eigenspinor with energy $E_n$ at site $i$. It is also possible to probe the magnetic texture in order to find out the nature of the magnetic ordering in the system. The primary signature of the topologically nontrivial phase consists of the zero-bias peak (ZBP) in the LDOS arising from the Majorana bound state localized at the end of the chain. In the topological phase the ZBP is isolated from other excitations by a minigap. The LDOS for the trivial gapped phase does not display a ZBP and should exhibit a robust gap everywhere in the wire, in stark contrast to the gapless phase for which the LDOS is non-vanishing near the Fermi level and does not exhibit a minigap. Preliminary experimental results on atomic chains include an observation of a ZBP at the chain ends and spatial modulation of magnetization observed using spin-polarized STM [15], both promising indications of the presence of Majorana states.

\begin{figure}
\includegraphics[width=0.90\columnwidth]{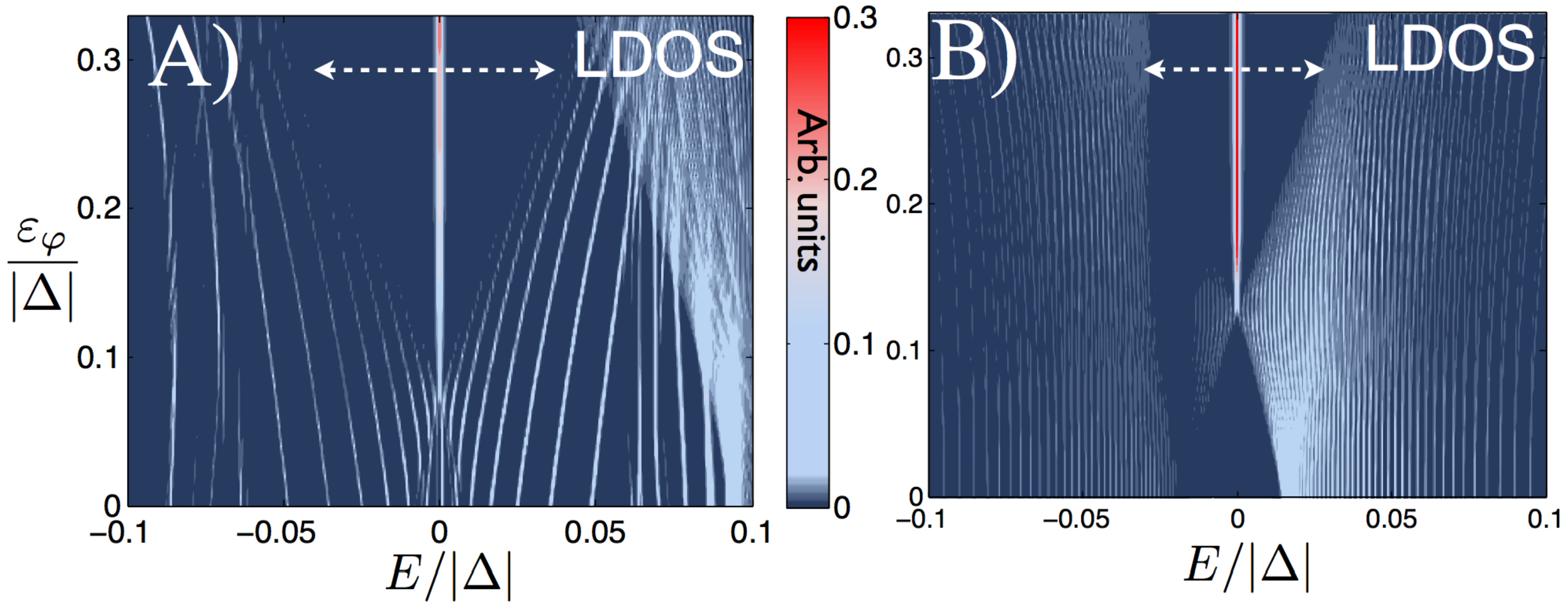}
\caption{ A): LDOS at the end of the chain of 100 atoms. The system is gapless in the absence of supercurrent and enters the topologically non-trivial gapped phase signalled by a ZBP and the opening of a minigap (white arrow). Parameters are $\theta=2\pi/5, k_ha=\pi/10, k_Fa=4.8\pi, \epsilon_0=0.05|\Delta|, \xi=50a$ . B):  Same as A) but for a system that is in the trivial gapped phase in the absence of supercurrent. The parameters are $\theta=\pi/5, k_ha=\pi/2, k_Fa=4\pi, \epsilon_0=0.015|\Delta|, \xi=50a$. }\label{fig4}
\end{figure}

The supercurrent-induced modifications to the phase diagram discussed above are observable in the LDOS. In Fig.~\ref{fig4} we plot the LDOS at the end of the chain. The parameters corresponding  Fig.~\ref{fig4}  A) places the helix in the gapless phase in the absence of supercurrent $\epsilon_\phi=0$. By increasing the phase gradient the system enters the topological phase with Majorana end states and the LDOS exhibits a clear ZBP. In Fig.~\ref{fig4}  B) the system is initially in the trivial state but undergoes a topological phase transition signalled by the appearance of a ZBP and the opening of a minigap. 

Our results have encouraging implications for the applications of topological superconductivity. If the microscopic parameters place a chain in the gapless or in the trivial gapped state, it could be possible to tune it to the topological phase in the case of a nonplanar helix.  In addition, the textures in Fig.~\ref{fig2} enable convenient switching between the nontrivial and trivial phases by supercurrent,  providing for example a means to braid Majorana states in "the Majorana necklace" \cite{li}. Instead of a rotating magnetic field one could carry out the braiding by rotating the direction of the supercurrent. A planar helix $\theta=\pi/2$ could be distorted by an external magnetic field to a non-planar configuration to render the  supercurrent control more effective.  Supercurrent provides a valuable control parameter in studying the topological phase transitions in experiments.  

\emph{Conclusion}-- In this work we studied the effects of supercurrent on the topological properties of Shiba chains and provided the phase diagram for an arbitrary helical magnetic texture. We discovered that for non-planar magnetic textures supercurrent control can be employed in tuning the system from the gapless phase to the nontrivial gapped phase. Supercurrent also enables switching the system between topological and trivial gapped phases, in some cases significantly increasing the nontrivial phase in the phase diagram.  For a planar magnetic helix supercurrent mostly drives the system towards the gapless phase. The LDOS, accessible in STM measurements, exhibits clear signatures of the predicted supercurrent-modified phase diagram.

The authors would like to thank Stevan Nadj-Perge, Alex Weststr\"om and Kim P\"oyh\"onen for discussions. T. O. acknowledges the Academy of Finland for support.

\appendix

\onecolumngrid
\section*{Supporting material}
\subsection*{Integrals}
Here we explain a number of technical results employed in the main text. We begin with expression (4) 
\begin{equation}\label{ap}
J_E(\vec{r}) =-JS\int\frac{d\vec{k}}{(2\pi)^3}e^{i\vec{k}\cdot r} \frac{(E+\frac{\vec{k}\cdot\nabla\varphi}{2m})+\tau_z\xi_k+|\Delta|\tau_x}{(E+\frac{\vec{k}\cdot\nabla\varphi}{2m})^2-\xi_k^2-|\Delta|^2}.
\tag{A1}
\end{equation}
for a finite $\vec{r}$ along the lines explained in Ref.~[12].  Compared to Ref.~[12], we have an added complication arising from the supercurrent-induced term proportional to $\nabla\phi$. Noting that $k_F$ is the characteristic magnitude of momentum, we can approximate $\vec{k}\cdot\nabla\varphi \approx \vec{k_F}\cdot\nabla\varphi$, with 
 $\vec{k}_F\parallel\vec{k}$ and $|\vec{k}_F|=k_F$. Defining $\tilde{\epsilon}_\phi=\frac{v_F|\nabla\phi|}{2}$, Eq.~(\ref{ap}) yields
\begin{equation}
J_E(\vec{r})=-JS\int\frac{d\vec{k}}{(2\pi)^3}e^{i\vec{k}\cdot r} \frac{E+\left(\frac{\vec{k}\cdot\nabla\varphi}{2m}\right)+\tau_z\xi_k+|\Delta|\tau_x}{E^2-\xi_k^2-|\Delta|^2}+\mathcal{O}(|\frac{\tilde{\epsilon}_\phi}{\Delta}|^2,\frac{E\tilde{\epsilon}_\phi}{|\Delta|^2} ).
\tag{A2}
\end{equation}
The corrections on the RHS arise from the $\nabla\phi$-dependent terms in the denominator on the RHS in Eq.~(\ref{ap}) that are omitted below. These corrections are always small in the considered parameter regime where $\frac{\tilde{\epsilon}_\phi}{|\Delta|}, \frac{E}{|\Delta|}\ll1$. Physically the condition $\frac{\tilde{\epsilon}_\phi}{|\Delta|}\ll1$ implies that we are considering phase gradients for which the corresponding supercurrent is much smaller than the critical current, while $\frac{E}{|\Delta|}\ll1$ is appropriate for deep impurities in the dilute limit, as explained below. Fixing the direction of $\mathbf{r}$, moving over to spherical coordinates, and letting
$\xi_k = \frac{k^2}{2m}-\mu \approx v_F(k-k_F)$ we obtain mostly the same integrals that are solved in Ref.~[12]. After noting that the integral of the azimuthal part of the scalar product of $\hat{k}$ with $\nabla\phi$ vanishes, the additional integral we need to solve is
\begin{equation}
\int\frac{d\vec{k}}{(2\pi)^3}\frac{e^{i\vec{k}\cdot \vec{r}}\left(\frac{\vec{k}\cdot\nabla\varphi}{2m}\right)}{E^2-\xi_k^2-|\Delta|^2}= \int\frac{d\vec{k}}{(2\pi)^3}\frac{e^{i\vec{k}\cdot \vec{r}}\cos(\theta)\epsilon_\varphi}{E^2-\xi_k^2-|\Delta|^2}
\tag{A3}
\end{equation}
where we have introduced $\epsilon_\varphi \equiv \tilde{\epsilon}_\varphi\cos(\beta)$, $\beta$ being the angle between $\mathbf{r}$ and $\nabla\varphi$. After evaluating the integrals, we obtain
\begin{equation}\label{finite}
J_E(\vec{r})=-\frac{\alpha}{\sqrt{|\Delta|^2-E^2}}\frac{e^{-\frac{\sqrt{|\Delta|^2-E^2}}{v_F} r}}{k_Fr}\left[ \left(E\sin k_Fr-i\epsilon_\varphi \cos k_Fr\right)\,I_{2\times2} +\sqrt{|\Delta|^2-E^2}\cos k_Fr\tau_z+|\Delta|\sin k_Fr\tau_x \right],
\tag{A4}
\end{equation}
where $\alpha\equiv\pi\nu_0SJ$. This is valid for distances $r > v_F\omega_D$, where $\omega_D$ is the cutoff energy scale [12].

For small $r$, we instead obtain
\begin{equation} \label{zero}
J_E(0)=-\frac{\alpha}{\sqrt{|\Delta|^2-E^2}}\left[ E\,I_{2\times2}\left(1 +\frac{1}{2}\frac{|\Delta|^2\epsilon_\varphi^2}{(|\Delta|^2-E^2)^2}\right)\, + |\Delta| \tau_x\left(1 + \frac{1}{6}\frac{(|\Delta|^2+2E^2)\epsilon_\varphi^2}{(|\Delta|^2-E^2)^2}\right) \right]+\mathcal{O} \left(\left|\frac{\epsilon_\varphi}{\Delta}\right|^4\right)
\tag{A5}
\end{equation}
which was obtained by setting $r = 0$ prior to integration.

\subsection*{Single-impurity problem}
The single-impurity eigenvalue problem, obtained from Eq.~(3) in the main text by setting $S_j=0$ for $i\neq j$, takes the form
\begin{equation}\label{zero1}
\left[1-(\mathbf{\hat{S}}_i\cdot \sigma)J_E(0)\right]\bar{\Psi}(\vec{r}_i)=0.
\tag{A6}
\end{equation}
This can be solved by inserting the expression (\ref{zero}) on the LHS. In the absence of supercurrent $\epsilon_\phi=0$ this equation has two well-known subgap solutions with energies $E=\pm \epsilon=\pm|\Delta|\frac{1-\alpha^2}{1+\alpha^2}$ where $\alpha=\pi\nu_0SJ$ as above.  The corresponding eigenstates are $\Psi_+(\vec{r}_i)\equiv|+i\rangle=|+\tau_x\rangle|\uparrow i\rangle$ and $\Psi_-(\vec{r}_i)\equiv|-i\rangle=|-\tau_x\rangle|\downarrow i\rangle$, where $\tau_x|\pm\tau_x\rangle=\pm|\pm\tau_x\rangle$ and  $\vec{\hat{S}}_i\cdot\sigma| \uparrow/ \downarrow i\rangle=\pm|\uparrow/ \downarrow i\rangle$.  The energies of deep impurities, for which $\alpha\sim 1$, simplify to $\pm\epsilon_0=\pm|\Delta|(1-\alpha)$.  Restoring finite $\epsilon_\phi$, the supercurrent-induced modification to the deep-impurity states can be straightforwardly solved, yielding $\pm\tilde{\epsilon}_0 =  \pm\epsilon_0\mp \frac{|\Delta|}{6}\left(\frac{\tilde{\epsilon}_\phi}{|\Delta|} \right)^2$.

\subsection*{Effective Hamiltonian for a chain}
The tight-binding model for multiple impurities can be derived from Eq.~(3) in the article [12]. Moving over the $i=j$ term from the sum to the LHS, we get
\begin{equation}
\left[1-(\mathbf{\hat{S}}_i\cdot \sigma)J_E(0)\right]\bar{\Psi}(\vec{r}_i)=-\sum_{j\neq i}(\mathbf{\hat{S}}_j\cdot \sigma)J_E(\vec{r}_i-\vec{r}_j)\bar{\Psi}(\vec{r}_j)
\tag{A6}
\end{equation}
Following Ref.~[12], we can make use of the fact that $(\mathbf{\hat{S}}_i\cdot \sigma)^2 = 1$ and multiply to get
\begin{equation} \label{chain}
\left[\mathbf{\hat{S}}_i \cdot \sigma-J_E(0)\right]\bar{\Psi}(\vec{r}_i)=-\sum_{j\neq i}(\mathbf{\hat{S}}_i\cdot \sigma)(\mathbf{\hat{S}}_j\cdot \sigma)J_E(\vec{r}_i-\vec{r}_j)\bar{\Psi}(\vec{r}_j).
\tag{A7}
\end{equation}
At this point we use the results we attained for $J_E(0)$ and $J_E(\vec{r})$. By denoting the distance between the magnetic moments by $a$, the RHS of Eq.~(\ref{chain}) is proportional to $\frac{1}{k_Fa}$, which satisfies $\frac{1}{k_Fa}\ll 1$ in the dilute impurity limit. This guarantees that a dilute chain of deep impurities satisfy $E\sim { \rm max} \{ \tilde{\epsilon}_0,\frac{|\Delta|}{k_Fa}\} \ll |\Delta|$.   We can set $E= 0$, $\alpha=1$ on the RHS, omitting small corrections  $\mathcal{O}(\frac{1-\alpha}{ k_Fa})$ [12]. Solving for E on the LHS, we get the equation
\begin{equation}\label{ap1}
E\bar{\Psi}(\vec{r}_i) = \left[|\Delta|\left(\mathbf{\hat{S}}_i \cdot \sigma-(\alpha+\frac{\epsilon_\phi^2}{6|\Delta|^2})\tau_x\right) -|\Delta| \displaystyle\sum_{j\neq i} (\mathbf{\hat{S}}_i\cdot \sigma)(\mathbf{\hat{S}}_j\cdot \sigma)J_0(\vec{r}_{ij})_{\vert{\alpha=1}}\right]\bar{\Psi}(\vec{r}_{ij})+\mathcal{O}(\frac{1}{k_Fa}\frac{\epsilon_0\epsilon_\phi}{|\Delta|^2}, \frac{1}{k_Fa}\frac{\epsilon_\phi^2}{|\Delta|^2}),
\tag{A8}
\end{equation} 
where we have indicated the supercurrent-dependent corrections that consist of products of at least three small parameters. Calculating matrix elements of Eq.~(\ref{ap1}) between decoupled states $|\pm i\rangle$, we can project the problem to the low-energy subspace and end up at Eq.~(5) and (6) in the main text.


\begin{thebibliography}{16}
\bibitem{qi}X.-L. Qi and S.-C. Zhang,  Rev. Mod. Phys. \textbf{83}, 1057 (2011).
\bibitem{schnyder} A. P. Schnyder, S. Ryu, A. Furusaki, and A. W. W. Ludwig, Phys. Rev. B \textbf{78}, 195125 (2008);  S. Ryu, A. P. Schnyder, A. Furusaki, and A. W. W. Ludwig, New J. Phys. \textbf{12}, 065010 (2010).
\bibitem{kitaev1} A. Y. Kitaev, Phys. Usp. \textbf{44}, 131 (2001).
\bibitem{nayak} C. Nayak, S. H. Simon, A. Stern, M. Freedman, and S. Das Sarma, Rev. Mod. Phys. \textbf{80}, 1083 (2008).
\bibitem{kitaev2} A. Y. Kitaev, Ann. Phys. \textbf{303}, 2 (2003).
\bibitem{kitaev3} A. Y. Kitaev, Ann. Phys. \textbf{321}, 2 (2006).
\bibitem{alicea2} J. Alicea, Y. Oreg, G. Refael, F. von Oppen, and M. P. A. Fisher, Nature Phys. \textbf{7}, 412 (2011).
\bibitem{choy} T. P. Choy, J. M. Edge, A. R. Akhmerov, and C. W. J. Beenakker, Phys. Rev. B \textbf{84}, 195442 (2011).
\bibitem{np} S. Nadj-Perge, I. K. Drozdov, B. A. Bernevig, and A. Yazdani, Phys. Rev. B \textbf{88}, 020407(R) (2013).
\bibitem{vazifeh} M.M. Vazifeh, M. Franz, Phys. Rev. Lett. \textbf{111}, 206802 (2013).
\bibitem{poyh}K. P\"oyh\"onen, A. Weststr\"om, J. R\"ontynen and T. Ojanen, Phys. Rev. B \textbf{89}, 115109 (2014).
\bibitem{pientka2} F. Pientka, L. I. Glazman and F. von Oppen, Phys. Rev. B \textbf{88}, 155420 (2013).
\bibitem{pientka3} F. Pientka, L. I. Glazman and F. von Oppen, Phys. Rev. B \textbf{89}, 180505 (2014).
\bibitem{heimes}  A. Heimes, P. Kotetes, G. Sch\"on, arXiv:1402.5901.
\bibitem{NP} S. Nadj-Perge, I. Drozdov, S. Jeon, J. Seo, B. A. Bernevig, A. Yazdani, APS March meeting 2014 abstract: Z46.00010: Experimental search for Majorana fermions in chains of magnetic atoms on a superconductor

\bibitem{braunecker1}B. Braunecker, G. I. Japaridze, J. Klinovaja, and D. Loss, Phys. Rev. B \textbf{82}, 045127 (2010).
\bibitem{klinovaja2}J. Klinovaja, M. J. Schmidt, B. Braunecker, and D. Loss, Phys. Rev. Lett. \textbf{106}, 156809 (2011).
\bibitem{klinovaja}J. Klinovaja, P. Stano, and D. Loss, Phys. Rev. Lett. \textbf{109}, 236801 (2012).
\bibitem{kjaergaard}M. Kjaergaard, K. W\"olms, and K. Flensberg, Phys. Rev. B \textbf{85}, 020503 (2012).
\bibitem{ojanen2}T. Ojanen, Phys. Rev. B \textbf{88}, 220502  (2013).
\bibitem{klinovaja3}  J. Klinovaja, P. Stano, A. Yazdani and D. Loss, Phys. Rev. Lett. \textbf{111}, 186805 (2013).
\bibitem{braun}  B. Braunecker and P. Simon,  Phys. Rev. Lett. \textbf{111}, 147202 (2013).

\bibitem{lutchyn}R. M. Lutchyn, J. D. Sau, and S. Das Sarma, Phys. Rev. Lett. \textbf{105}, 077001 (2010).
\bibitem{oreg}Y. Oreg, G. Refael, and F. von Oppen, Phys. Rev. Lett. \textbf{105}, 177002 (2010).
\bibitem{mourik}V. Mourik, K. Zuo, S. M. Frolov, S. R. Plissard, E. P. A. M. Bakkers, and L. P. Kouwenhoven, Science \textbf{336}, 6084 (2012).
\bibitem{das} A. Das, Y. Ronen, Y. Most, Y. Oreg, M. Heiblum, and H Shtrikman, Nat. Phys. \textbf{8}, 887 (2012).
\bibitem{menzel}M. Menzel, Y. Mokrousov, R. Wieser, J. E. Bickel, E. Vedmedenko, S. Bl\"ugel, S. Heinze, K. von Bergmann, A. Kubetzka, and R. Wiesendanger, Phys. Rev. Lett. \textbf{108}, 197204 (2012).

\bibitem{romito} A. Romito, J. Alicea, G. Refael, F. von Oppen, Phys. Rev. B \textbf{85}, 020502(R) (2012).
\bibitem{on} See the supporting material. 
\bibitem{yu} L. Yu, Acta Phys. Sin.  \textbf{21}, 75 (1965).
\bibitem{shiba} H. Shiba, Prog. Theor. Phys.  \textbf{40}, 435 (1968).
\bibitem{rusinov} A. I. Rusinov, JETP Lett.  \textbf{9}, 85 (1969).

\bibitem{li}  J. Li, T. Neupert, B. A. Bernevig and  A. Yazdani, arXiv:1404.4058

\end{thebibliography}
\end{document}